%% file: draft_pi0pi0psip.tex
\RequirePackage{lineno}
\documentclass[twocolumn,showpacs,superscriptaddress,amsmath,amssymb]{revtex4-1}
\usepackage{graphicx}
\usepackage{epsfig}
\usepackage{dcolumn}
\usepackage{bm}
\usepackage{overpic}
\usepackage{color}
\usepackage{xspace}

\usepackage{subfigure}
\usepackage{url}
\usepackage{hyperref}

\include{def-com}

\def \jpsi {J/\psi}
\def \psip {\psi(3686)}
\def \epem {e^+e^-}
\def \lplm {\ell^+\ell^-}
\def \pzpz {\pi^0\pi^0}
\def \pppm {\pi^{+}\pi^{-}}

\def \zcbz {Z_c(3900)^0}
\def \piz  {\pi^0}

\def \gev  {\mbox{GeV}\xspace}

\def \gevcc{\mbox{GeV/$c^2$}\xspace}
\def \mev  {\mbox{MeV}\xspace}
\def \mevcc{\mbox{MeV/$c^2$}\xspace}
\def \ifb  {\mbox{fb$^{-1}$}\xspace}

\begin{document}
\title{\boldmath Measurement of $\epem \to \pzpz\psip$ at $\sqrt{s}$ from 4.009 to 4.600 $\gev$
                 and observation of a neutral charmoniumlike structure}
\input{authors_apr2015}
\date{\today}
\begin{abstract}
  Using $\epem$ collision data collected with the BESIII detector at the BEPCII collider corresponding to an integrated luminosity of 5.2~$\ifb$ at center-of-mass energies ($\sqrt{s}$) from 4.009 to 4.600~$\gev$, the process $\epem\to\pzpz\psip$ is studied for the first time. The corresponding Born cross sections are measured and found to be half of those of the reaction $\epem\to\pppm\psip$. This is consistent with the expectation from isospin symmetry. Furthermore, the Dalitz plots for $\pzpz\psip$ are accordant with those of $\pppm\psip$  at all energy points, and a neutral analogue to the structure in $\pi^\pm \psip$ around 4040~$\mevcc$ first observed at $\sqrt{s}=4.416~\gev$ is observed in the isospin neutral mode at the same energy.
\end{abstract}

\pacs{14.40.Rt, 14.40.Pq, 13.66.Bc}

\maketitle

The vector charmoniumlike state $Y(4360)$ was observed and subsequently confirmed in $\epem \to (\gamma_{\rm ISR}) \pppm \psip$ by BaBar, Belle, and BESIII~\cite{Y4360BaBar, Y4360Belle, Y4360Bes}, where $\gamma_{\rm ISR}$ refers to an Initial State Radiation (\textsc{ISR}) photon.
However, the nature of the $Y(4360)$ remains mysterious~\cite{NoConvention}, as is the case for other states of the $Y$ family, {\it e.g.} the $Y(4260)$ observed in $\epem\to(\gamma_{ISR})\pppm\jpsi$~\cite{Y4260BaBar, Y4260Cleo, Y4260Bell, Y4260Bes}.
Many theoretical interpretations have been proposed to explain the underlying structure of the $Y$ family of states~\cite{hybrid,tetraquark,molecule}.
It is therefore compelling to study the $Y(4360)$ in its $\pzpz$ transition to $\psip$ and to examine isospin symmetry.

In recent years, a new pattern of charmoniumlike states, the $Z_c^{\pm}$'s, was observed in the systems of a charged pion and a low-mass charmonium state~\cite{Y4260Bell,ZC3900BES,ZC3900CLEO,ZC4020BES,Y4360Bes}, as well as in charmed mesons pairs~\cite{ZC3885BES,ZC3885BES2,ZC4025BES}.
The observation of $Z_c^{\pm}$ particles and of similar states in the bottomonium system~\cite{zb} indicates the discovery of a new class of hadrons~\cite{exotics}.
More recently, neutral charmoniumlike states, which are referred to as $Z_c^{0}$'s, have been reported in analogous
systems~\cite{4020zchc,4020zcDpi0,4020zcDDs,4020zcjpsipi0}. These are regarded as the neutral isospin partners of the $Z_c^{\pm}$'s.
A charmoniumlike structure observed in $\epem\to\pppm\psip$ by BESIII~\cite{Y4360Bes} was also reported
in Belle's latest updated result~\cite{Y4360Belle}.
By analogy, it is interesting to search for its neutral isospin partner in $\epem\to\pzpz\psip$.

In this Letter, we present a study of the process $\epem \to \pzpz$ $\psip$ at center-of-mass (c.m.)~energies ($\sqrt{s}$) from 4.009 to 4.600 $\gev$.
The corresponding Born cross sections are measured for the first time.
A new neutral structure is observed in the $\piz\psip$ invariant mass spectra around 4040~$\mevcc$.
The data samples used in this analysis were collected with the BESIII detector at 16 different c.m.~energies with a total integrated luminosity of 5.2 $\ifb$~\cite{luminosity}.
The c.m.~energies have been measured with di-muon events for each energy point~\cite{energycorr}.

Details on the features and capabilities of the BEPCII collider and the BESIII detector can be found in Ref.~\cite{besint}.
The {\sc geant4}-based~\cite{geant4} Monte Carlo (MC) simulation software package BOOST~\cite{Deng} is used to generate MC samples.
Simulated MC samples for the signal process $\epem\to\pzpz\psip$ with $\psip\to\pppm\jpsi$, $\jpsi\to\lplm$, and $\ell$ = $e$/$\mu$ (referred to as $\epem\to\pzpz\psip$ throughout this Letter)
and the background process $\epem\to\pppm\psip$ with $\psip \to \pzpz \jpsi$, and $\jpsi\to\lplm$ (referred to as $\epem\to\pppm\psip$ throughout this Letter) are generated at each c.m.~energy.
The $\epem$ collision is simulated with the {\sc{KKMC}}~\cite{kkmc1} generator incorporating the beam energy spread and \textsc{ISR}, where the cross section line shapes of $\epem \to \pzpz \psip$ and $\epem\to \pppm \psip$ are assumed to be the same and are taken from the latest results from Belle~\cite{Y4360Belle}.
The processes $\epem\to\pi^{0/+}\pi^{0/-} \psip$ and $\psip \to
\pi^{+/0} \pi^{-/0}\jpsi$ are simulated with the \textsc{Jpipi}
model~\cite{jpipi} of \textsc{Evtgen}~\cite{EVTGEN}.
As in Refs.~\cite{ZC4020BES,Y4360Bes}, the inclusive MC samples at $\sqrt{s}= 4.258$ and 4.358~$\gev$
are used to study the potential backgrounds.

The signal candidates  are required to have four charged tracks with zero net charge and at least four photon candidates.
The selection criteria for good charged tracks and photons, the separation between pions, electrons and muons, as well as the hit number required in the muon system for the $\mu^{+}\mu^{-}$ pair are the same as those in Ref.~\cite{Y4360Bes}.

A four-constraint (4C) kinematic fit imposing energy-momentum conservation under the hypothesis $\epem\to\gamma\gamma\gamma\gamma \pppm \lplm$ is performed, and $\chi^{2}_{4C} < 120$ is required.
For the events with more than four photons, the combination with the  least $\chi^{2}_{4C}$ is retained.
The pairing of photons into the two $\piz$ is chosen by minimizing $(M(\gamma_1\gamma_2) - M(\piz))^{2} + (M(\gamma_3\gamma_4) - M(\piz))^{2}$. The $\jpsi$ and $\piz$ candidates are selected by requiring $3.05 <M(\lplm)< 3.15$~$\gevcc$ and $|M(\gamma_i\gamma_j) - M(\piz)|< 20 $~\mevcc, where $M(\piz)$ is the $\pi^{0}$ mass according to the PDG~\cite{PDG}.
A seven-constraint (7C) kinematic fit with additional constraints on the two $\piz$ and $\jpsi$ masses~\cite{PDG} is imposed to suppress the non-$\piz\piz\pppm\jpsi$  backgrounds and improve the mass resolution.

Figure~\ref{pipirecoil} shows the scatter plots of the $\pppm$ recoil mass $M_{\rm recoil}(\pppm)$ versus $M(\pppm\jpsi)$ and the $M(\pppm\jpsi)$ spectra for the data samples at $\sqrt{s}= 4.226$, 4.258, 4.358, 4.416, and 4.600 $\gev$, which have relatively large statistics.
Vertical and horizontal bands at the $\psip$ mass position are observed clearly in the scatter plots, corresponding to the processes $\epem\to\pzpz$ $\psip$ and $\epem\to\pppm\psip$, respectively. The narrow peaks in the $M(\pppm\jpsi)$ spectra indicate the signal process $\epem\to\pzpz \psip$, while the relative broad bumps with position depending upon c.m.~energy are from $\epem\to\pppm\psip$.

\begin{figure*}[htbp]
 \centering
 \begin{overpic}[width=0.195\textwidth]{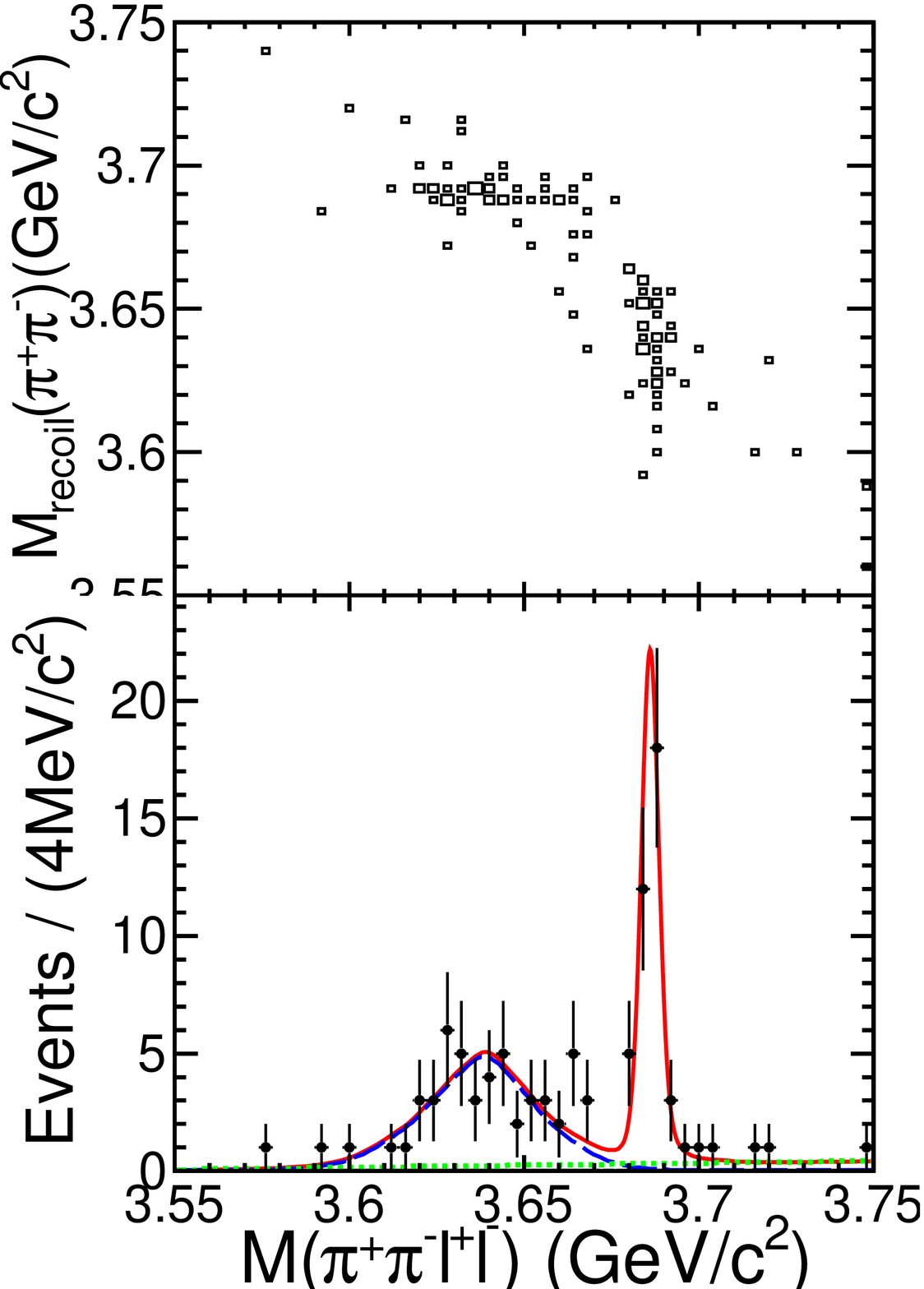}
 \put(47,90){ (a)}
 \end{overpic}
 \begin{overpic}[width=0.195\textwidth]{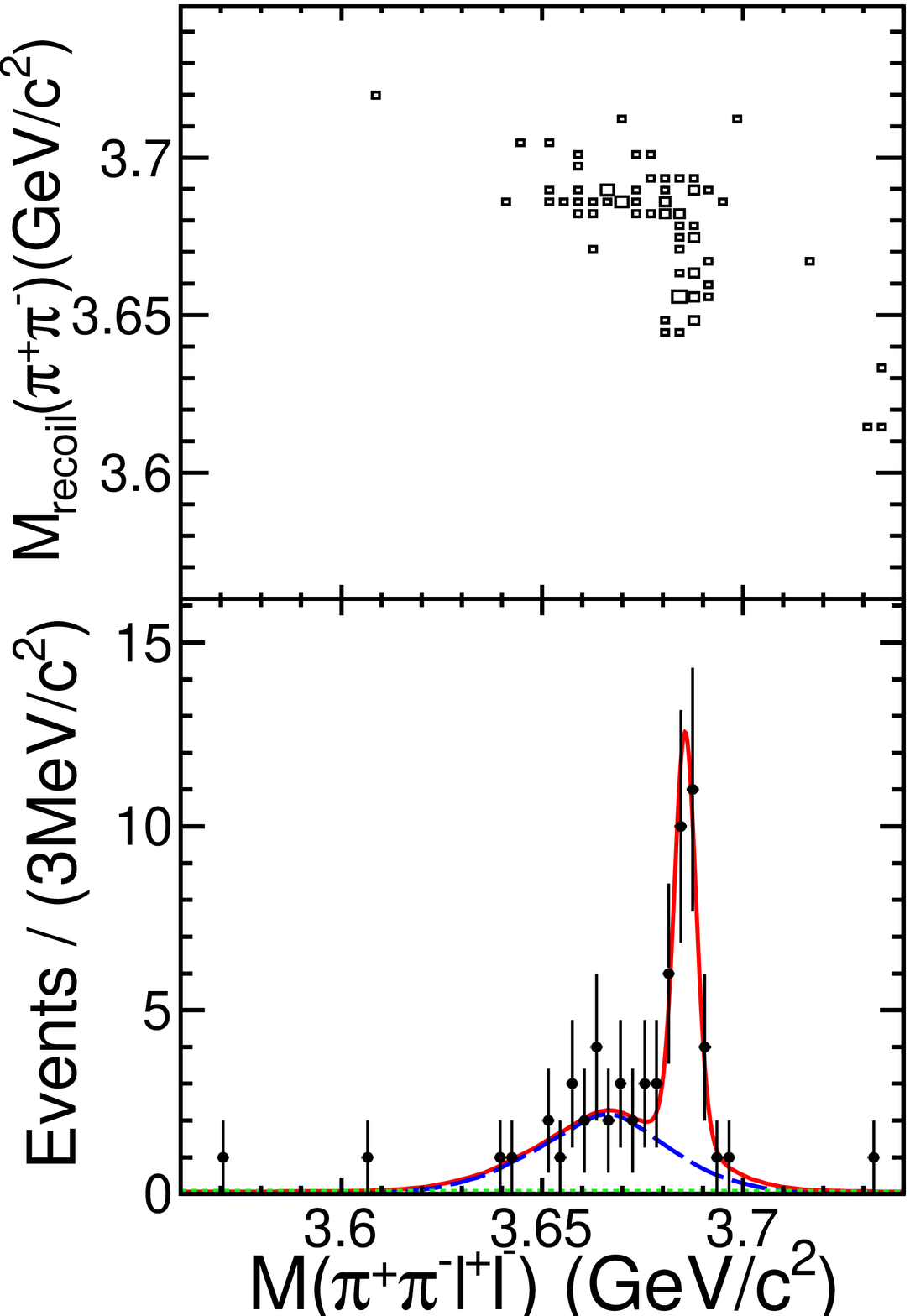}
 \put(47,90){ (b)}
 \end{overpic}
 \begin{overpic}[width=0.195\textwidth]{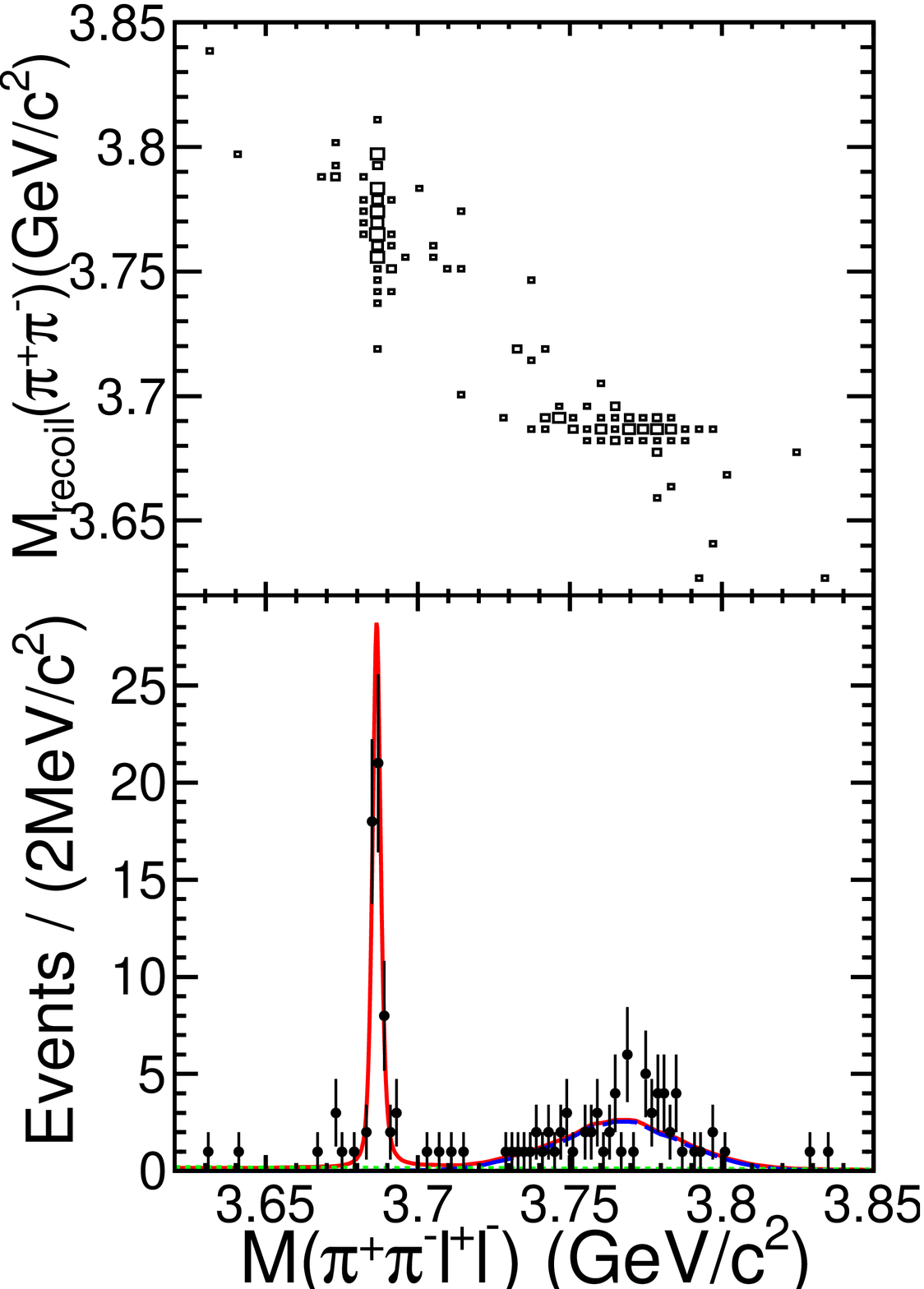}
 \put(47,90){ (c)}
 \end{overpic}
 \begin{overpic}[width=0.195\textwidth]{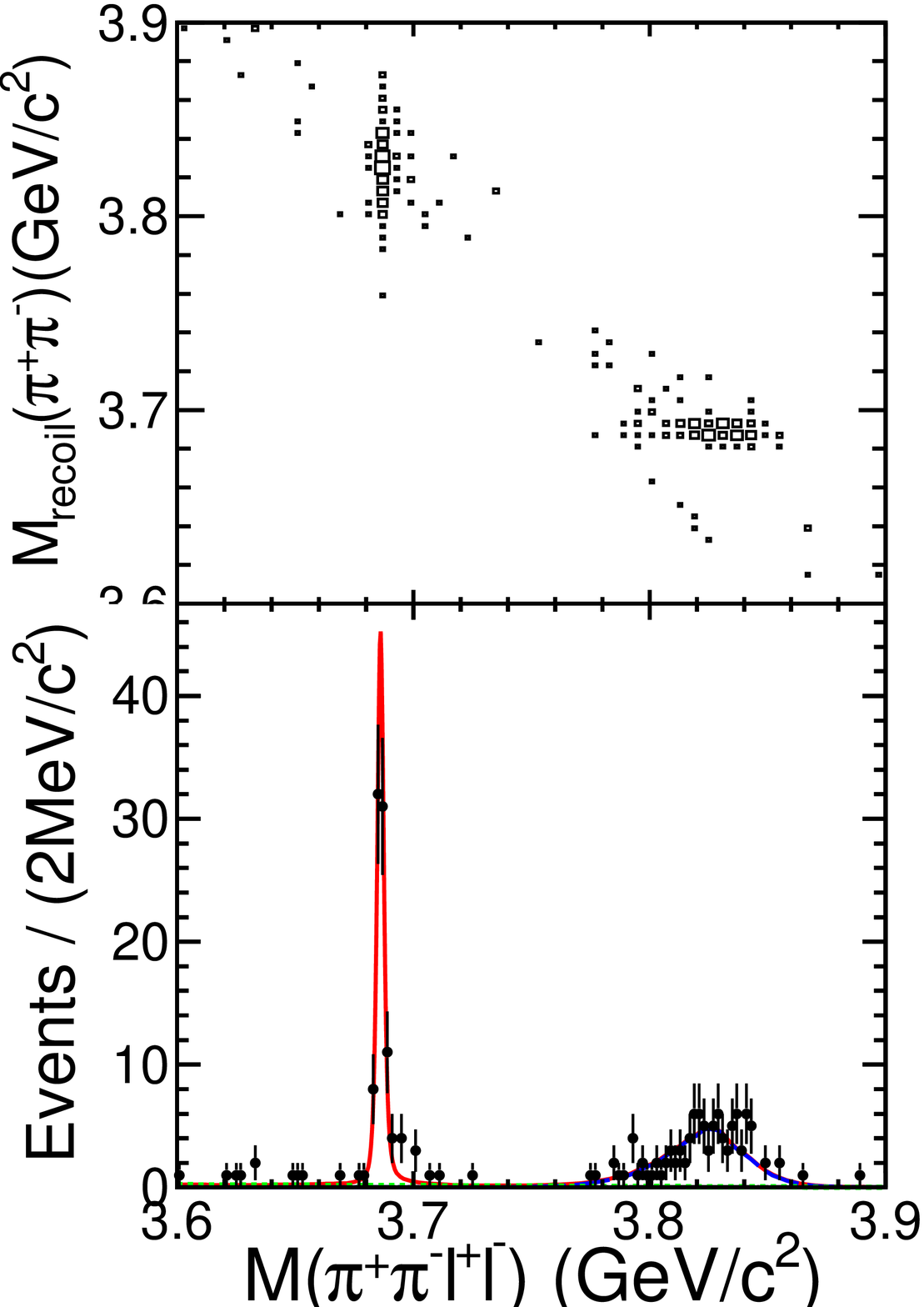}
 \put(47,90){ (d)}
 \end{overpic}
 \begin{overpic}[width=0.195\textwidth]{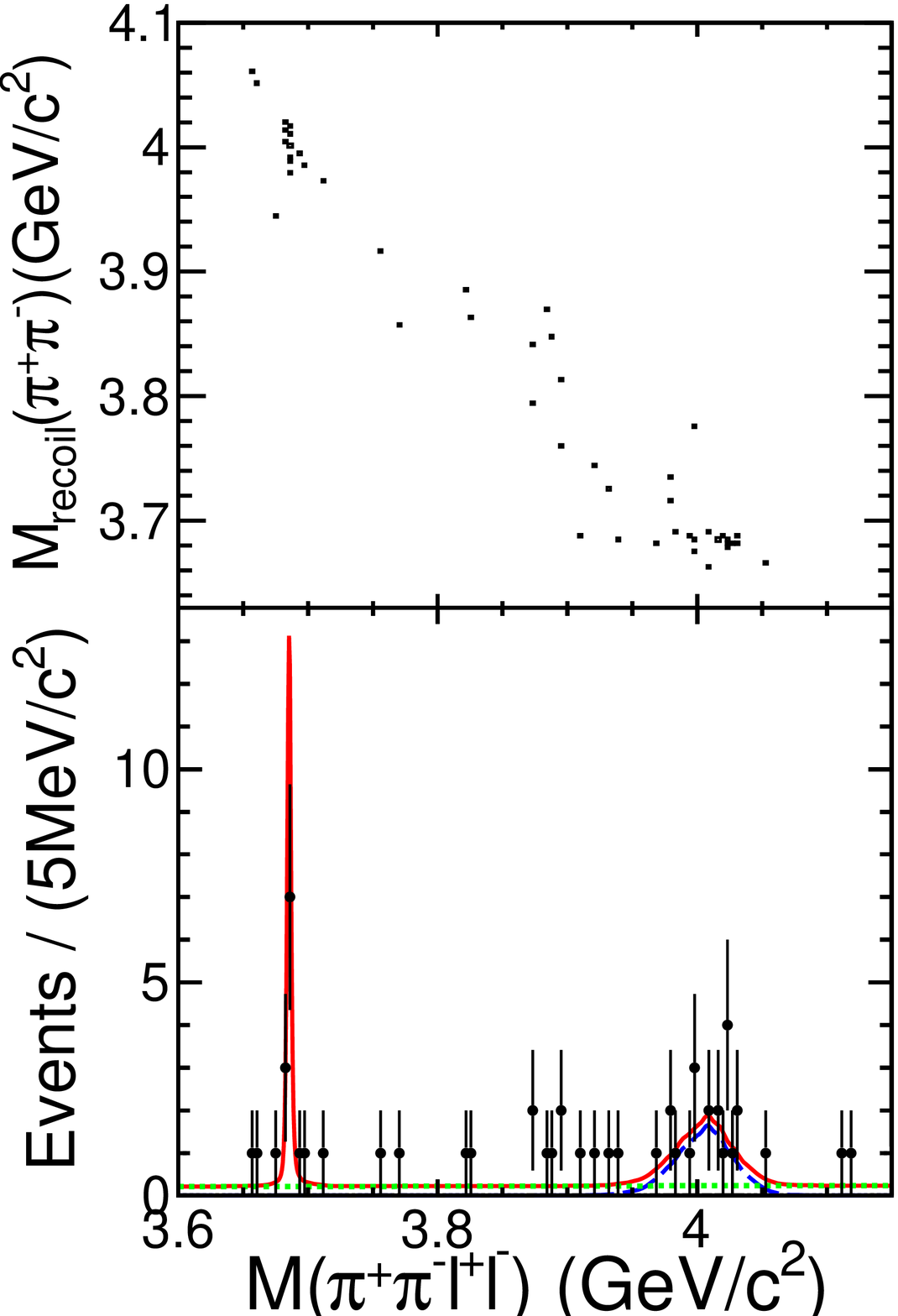}
 \put(47,90){ (e)}
 \end{overpic}
 \setlength{\belowcaptionskip}{-0.2cm}
 \caption{
  (Color online) Scatter plots of $M_{\rm recoil}(\pppm)$ versus $M(\pppm l^{+}l^{-})$ (top) and the $M(\pppm l^{+}l^{-})$ spectra (bottom).
  Dots with error bars are data; the solid curves show the results of the best fits; the dashed (blue) curves for the background $\epem\to\pppm\psip$; the short dashed (green) curves for the other backgrounds.
  The different columns show data at $\sqrt{s} = $ (a) 4.226, (b) 4.258, (c) 4.358, (d) 4.416, (e) 4.600~$\gev$, respectively.
        }
 \label{pipirecoil}
\end{figure*}

The inclusive and exclusive MC samples, as well as the data in the $\jpsi$ sideband region (selected by applying a 6C kinematic fit without the $\jpsi$ mass constraint instead of the 7C kinematic fit), are used to investigate the backgrounds.
The dominant background is $\epem\to\pppm\psip$, which has the same final states as the signal.
An unbinned maximum likelihood fit is performed to the $M(\pppm\jpsi)$ spectra to determine the signal yields.
In the fit, the probability density function (PDF) of $\epem\to$ $\pzpz\psip$ and $\epem \to \pppm \psip$ are described with the MC simulated shapes convolved with a Gaussian function, where the parameters of the Gaussian function are determined in the fit, in order to account for the resolution difference and potential mass shift between the data and MC simulation.
The other backgrounds are described with a linear function.
The fits curves are shown in Fig.~\ref{pipirecoil}.

The Born cross section is calculated from
\begin{equation}
\sigma^{\rm B}=\frac{N^{\rm obs}}{\mathcal{L}_{\rm int}~(1+\delta^{\rm r})~(1+\delta^{\rm v})~\mathcal{B}~\epsilon},
\label{equation}
\end{equation}
where $\mathcal{L}_{\rm int}$ is the integrated luminosity, $N^{\rm obs}$ is the signal yield from the fit,
$(1+\delta^{\rm r})$ is the ISR correction factor which is obtained by using a QED calculation~\cite{ISR} and incorporating the input lineshape
of the cross section which is taken to be the same as that of $\epem\to\pppm\psip$ from the Belle experiment~\cite{Y4360Belle},
$(1+\delta^{\rm v})$ is the vacuum polarization factor taken from a QED calculation with an accuracy of 0.5\%~\cite{VR},
$\mathcal{B}$ is the product of the branching fractions in the decay chain, 3.89\%, taken from the PDG~\cite{PDG}, and $\epsilon$ is the detection efficiency.
The numbers used in the Born cross section calculation and the cross sections are summarized in the Supplemental Material~\cite{supp}.
The comparison of the Born cross section of $\epem\to\pzpz\psip$ to that of $\epem\to\pppm\psip$ for the data samples with large luminosities is shown in Fig.~\ref{crosssection}.
The Born cross sections of $\epem \to\pppm\psip$ are also calculated with the corresponding event yields, and are consistent with the results in Ref.~\cite{Y4360Bes}. An alternative fit to the $M_{\rm recoil}(\pppm)$ spectra, which have a narrow peak for $\epem$ $\to\pppm\psip$ and a broad bump depending on c.m.~energy for $\epem\to\pzpz\psip$, is performed. In the fit, the PDF is described by the similar strategy with the $M(\pppm\jpsi)$ spectra. The resulting Born cross sections are consistent with the nominal values for the two processes.

  \begin{figure}[htb]
  \vskip  -0.3cm
  \hskip -0.0cm
  \begin{overpic}[width=0.35\textwidth]{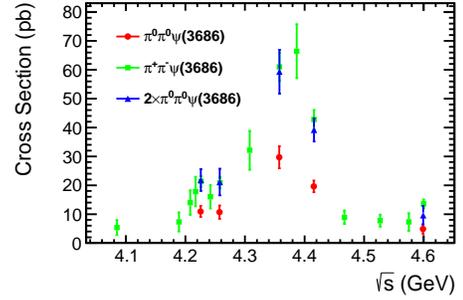}
  \end{overpic}
   \setlength{\belowcaptionskip}{-0.2cm}
  \caption{Born cross section of $\epem\to\pzpz\psip$ at $\sqrt{s}$ = 4.226, 4.258, 4.358, 4.416, 4.600~$\gev$, respectively. The dots (red) are the results obtained in this analysis, the triangles (green) are twice of our results and squares (blue) are the Born cross section of $\epem\to\pppm\psip$ from BESIII.}\label{crosssection}
  \end{figure}

For the data samples with small luminosities, only a small number of events survives.
The events within $3.676<M_{\rm recoil}(\pppm)<3.696~\gevcc$ are removed to suppress backgrounds from $\epem\to\pppm\psip$.
Upper limits at the 90\% confidence level (C.L.) on the Born cross sections are determined by using a frequentist method with profile likelihood treatment of systematic uncertainties~\cite{Rolke}.
The number of signal events ($N^{\rm obs}$) is counted in the region $3.671<M(\pppm\jpsi)<3.701~\gevcc$, while the number of background events ($N^{\rm bkg}$) is evaluated in the region $3.630<M(\pppm\jpsi)<3.660~\gevcc$ or $3.712<M(\pppm\jpsi)<3.742~\gevcc$.
In the calculation, the observed events are assumed to have a Poisson distribution and the event selection efficiencies to follow a Gaussian distribution.
The upper limits are shown in the supplemental material~\cite{supp}.

The cross section ratios, $R_{\pi^{+}\pi^{-}\psi(3686)}=\frac{\sigma(\epem\to\pzpz\psip)}{\sigma(\epem\to\pppm\psip)}$, are calculated for data samples with large luminosities and are listed in the Supplemental Material~\cite{supp},
where $\sigma(\epem\to\pppm\psip)$ are taken from Ref.~\cite{Y4360Bes}.
A set of common systematic uncertainties among the two processes, including those on luminosity, tracking efficiencies, and the requirements on the lepton tracks, cancel in the calculation. The weighted average of the ratios at $\sqrt{s} = 4.226$, 4.258, 4.358, 4.416~$\gev$ is $0.48 \pm 0.04 \pm 0.02$. Within uncertainties, the resulting $R_{\pi^{+}\pi^{-}\psi(3686)}$ is consistent with the value 0.5 expected from isospin symmetry.

The following sources of systematic uncertainty are considered in the cross section measurements.
The uncertainty on the efficiency for charged tracks (photons) is 1\% per track (photon)~\cite{trackerror,photonerror}.
The uncertainty on the hit number requirement in the muon counter is 4.2\%, obtained by studying a sample of $\epem$ $\to\pppm \jpsi$ events.
The uncertainty related with the kinematic fit is estimated with the same method as in Ref.~\cite{helixsys}, and is in the range 0.3\% to 1.2\% depending on c.m.\ energy.
The uncertainties of the  $\piz$ and $\jpsi$ invariant mass requirements are evaluated by tuning the corresponding MC distributions according to data, and are in the range 0.2\% to 0.5\% and 0.1\% to 0.3\%, respectively, depending on c.m.\ energy.
The uncertainties related to the fit procedure are investigated by varying the fit range, replacing the linear function for the background by a second-order polynomial function for background, and varying the width of the Gaussian function for the signal, and are in the range 1.6\% to 7.3\% depending on c.m.\ energy.
For the data samples with large luminosity, the detection efficiencies are estimated with the MC samples re-weighted according to the Dalitz plots distributions of $M^2(\piz\psip)$ versus $M^2(\pzpz)$ found in data.
The corresponding uncertainty is estimated by varying the weighting factors according to the statistical uncertainty in each bin.
For the data samples with low luminosity, the detection efficiencies are estimated with the {\sc Jpipi} model MC samples.  The corresponding systematic uncertainties are estimated with the data samples with large luminosity by comparing the efficiencies derived from the {\sc Jpipi} model MC sample with the nominal model.
The uncertainty associated with the ISR correction factor is studied by replacing the input cross section line shape with the latest results from BaBar~\cite{Y4360BaBar} in the KKMC generator, and is in the range 0.3\% to 2.4\% depending on c.m.\ energy.
The uncertainty of the vacuum polarization factor is 0.5\% from a QCD approach~\cite{VR}.

The uncertainty of the integrated luminosity is 1\%, as determined with large-angle Bhabha events~\cite{luminosity}.
The uncertainties of the branching fractions of the intermediate states are taken from the PDG~\cite{PDG}.
A summary of all considered systematic uncertainties is shown in the supplemental material~\cite{supp}.
Assuming all sources of systematic errors are independent,
the total uncertainties are the quadratic sums of the individual values, ranging from 7.8\% to 10.8\%, depending on the c.m.~energy.

\begin{figure*}[htbp]
 \centering
 \begin{overpic}[width=0.23\textwidth]{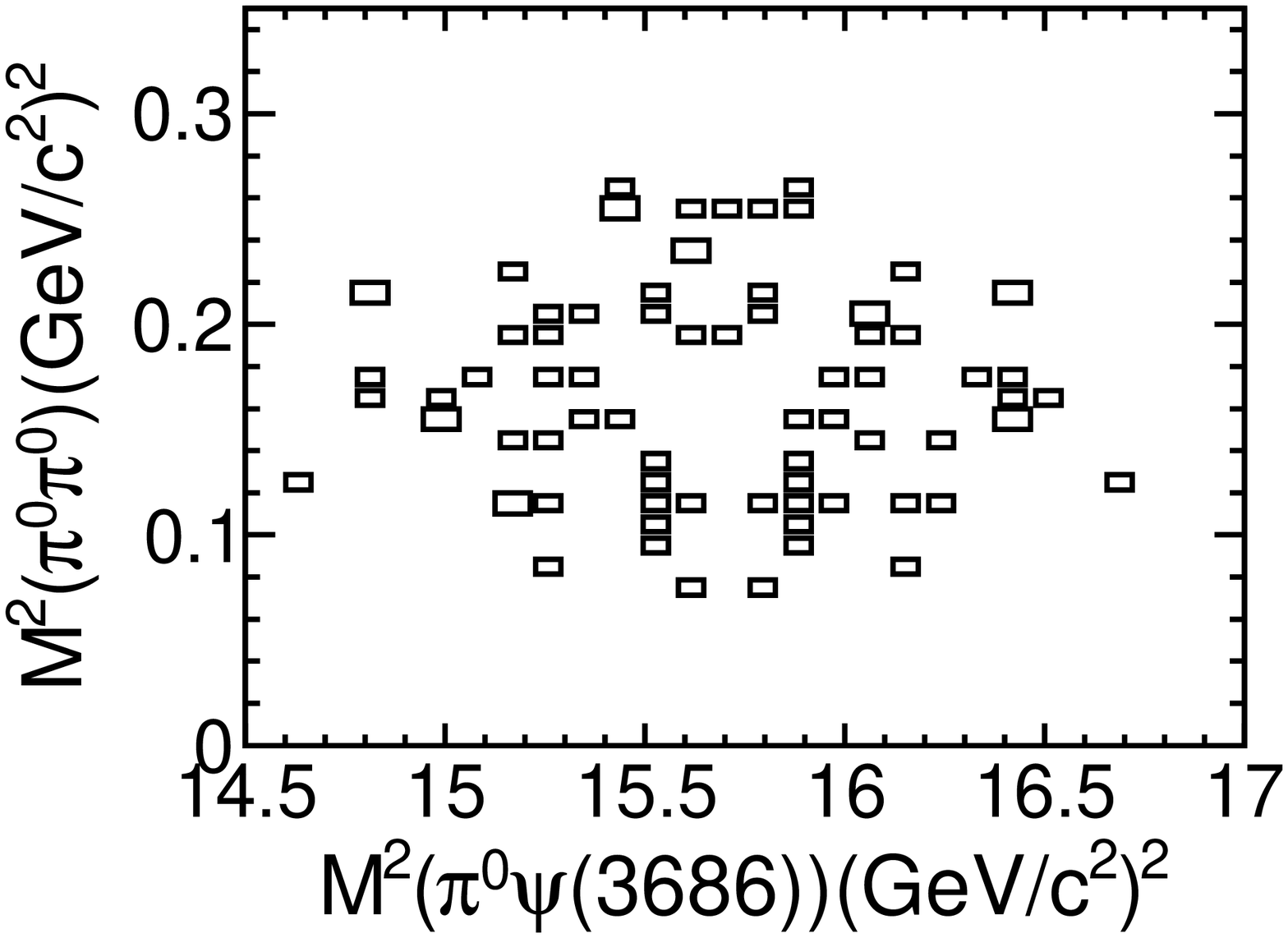}
 \put(77,62){ (a)}
 \end{overpic}
 \begin{overpic}[width=0.23\textwidth]{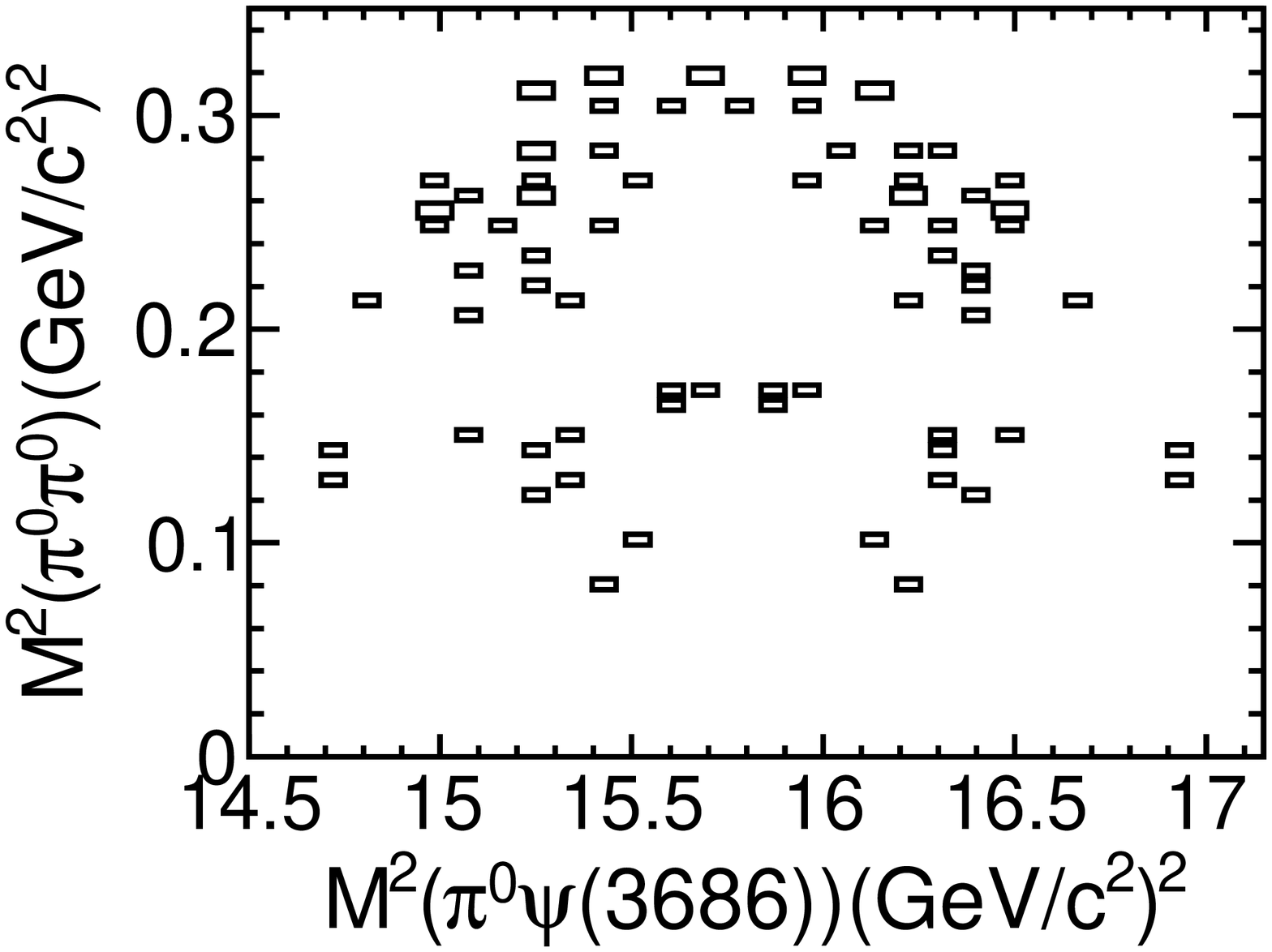}
 \put(77,62){ (b)}
 \end{overpic}
 \begin{overpic}[width=0.23\textwidth]{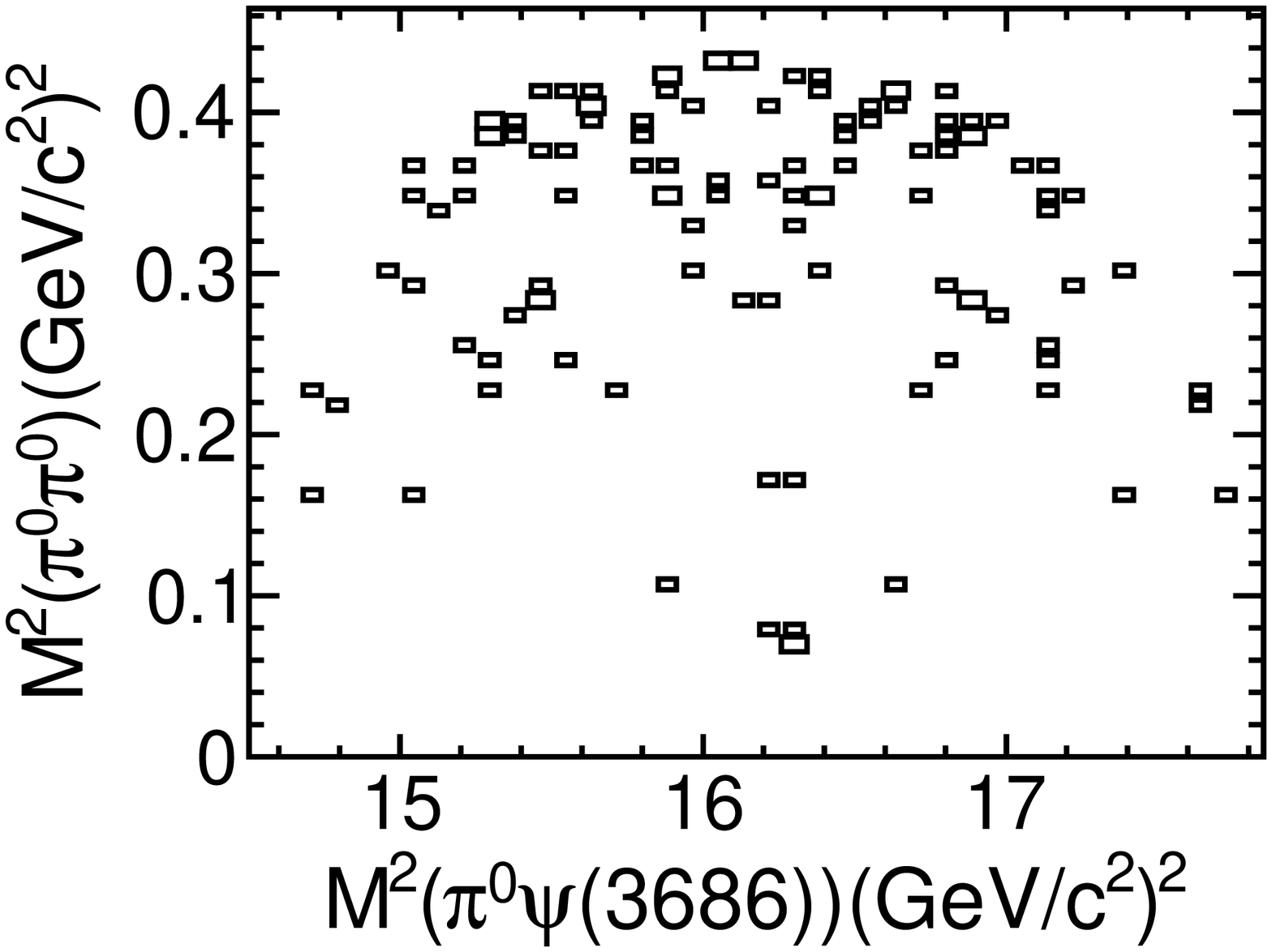}
  \put(77,62){ (c)}
 \end{overpic}
 \begin{overpic}[width=0.23\textwidth]{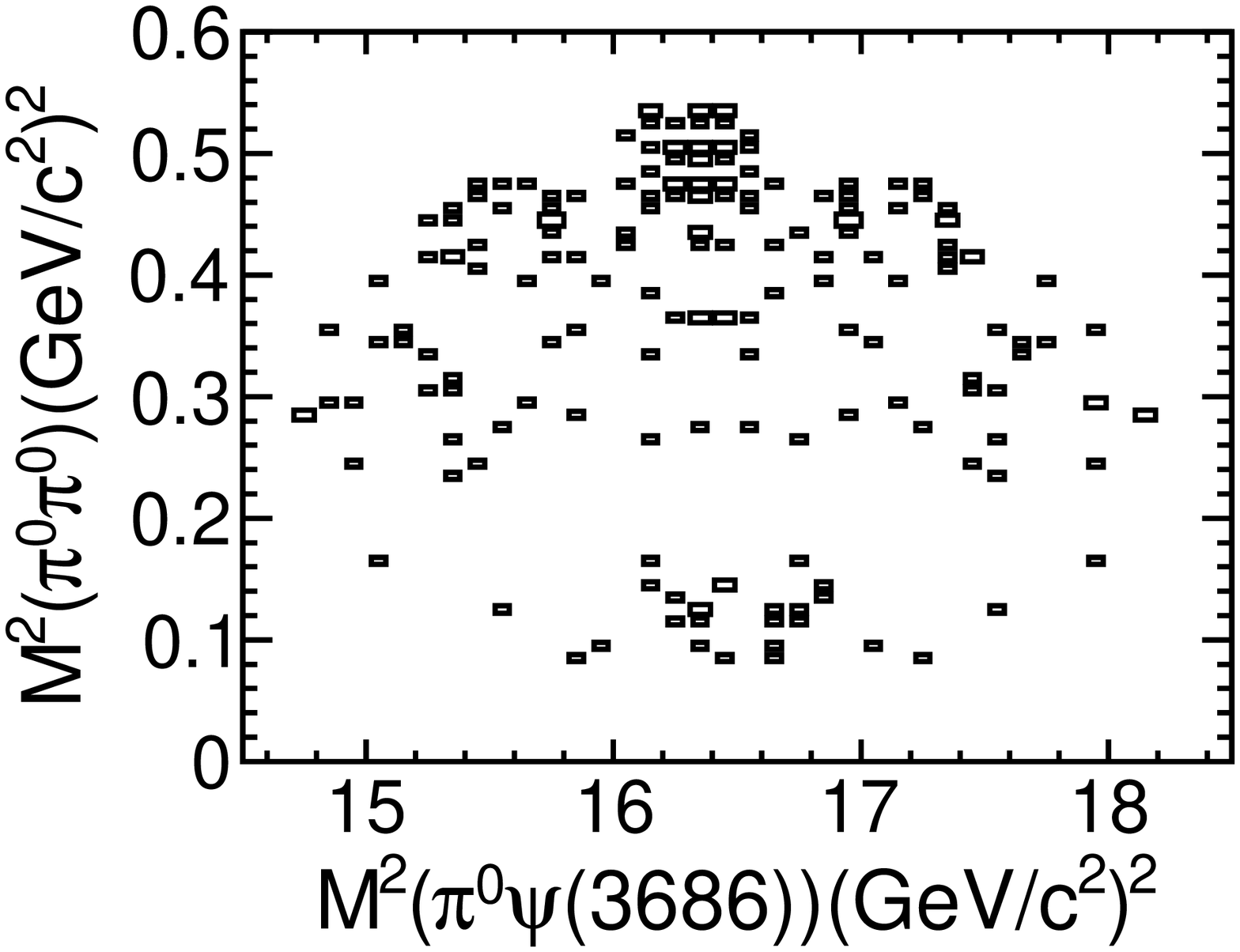}
 \put(77,62){ (d)}
 \end{overpic}
 \begin{overpic}[width=0.23\textwidth]{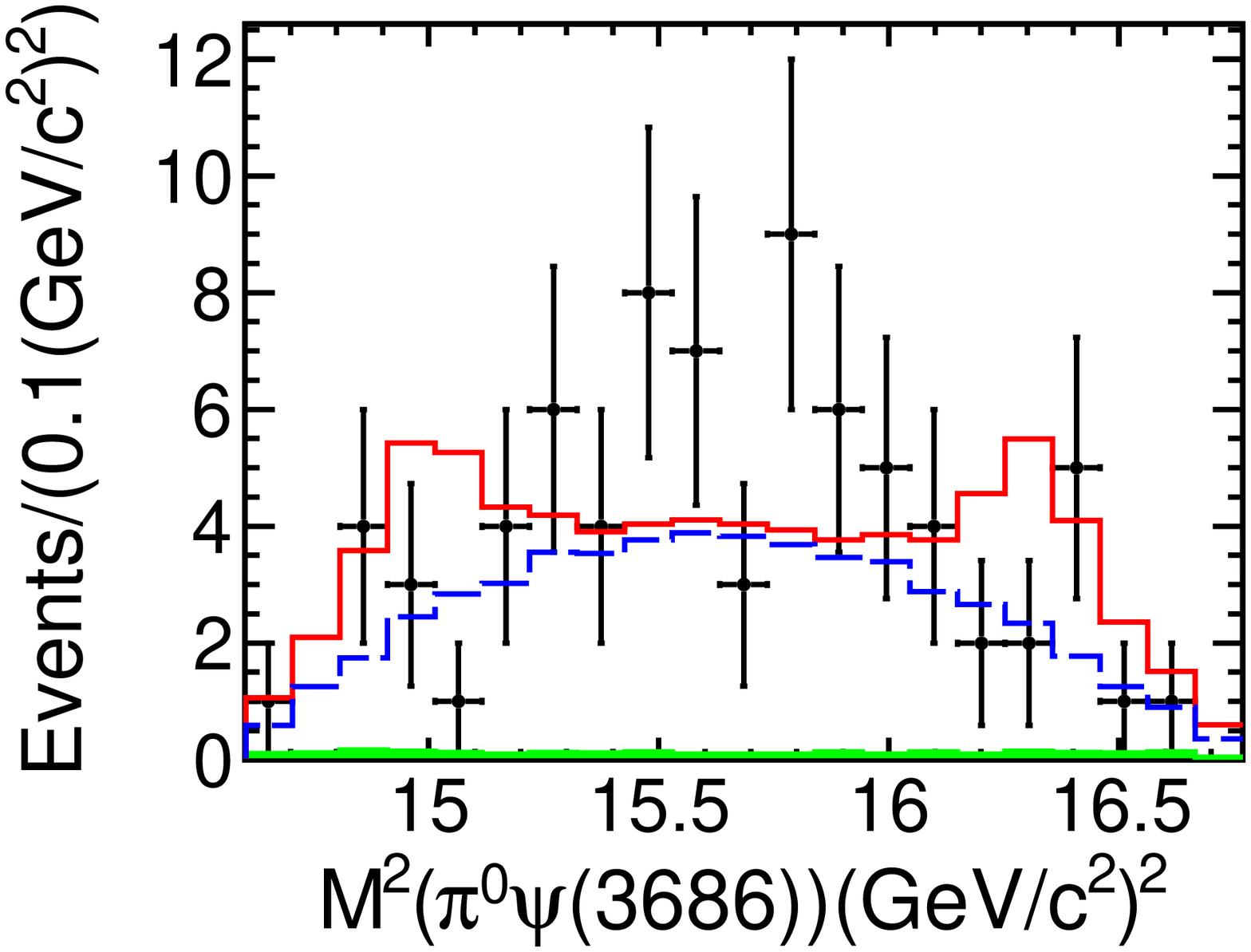}
 \end{overpic}
 \begin{overpic}[width=0.23\textwidth]{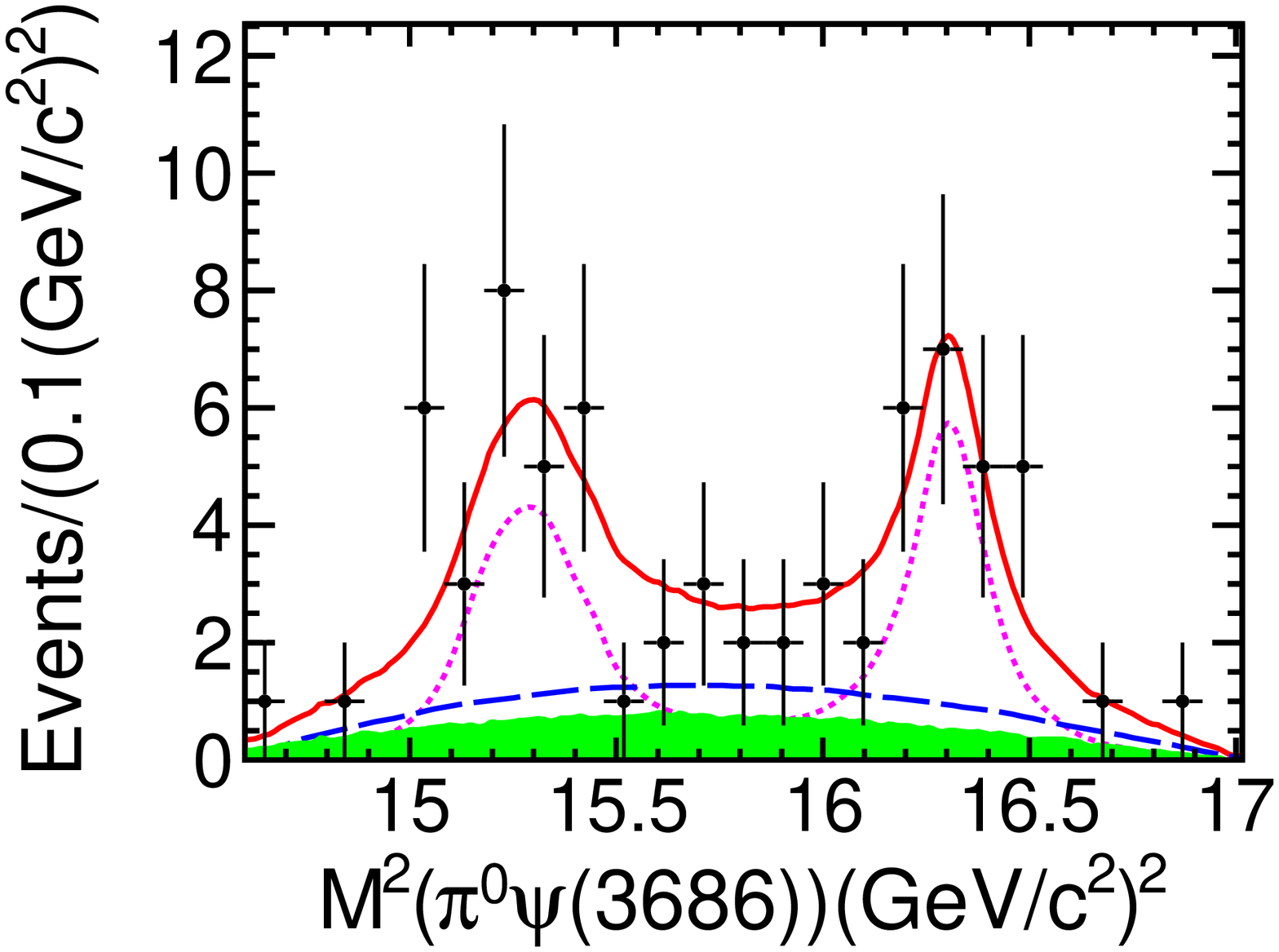}
 \end{overpic}
 \begin{overpic}[width=0.23\textwidth]{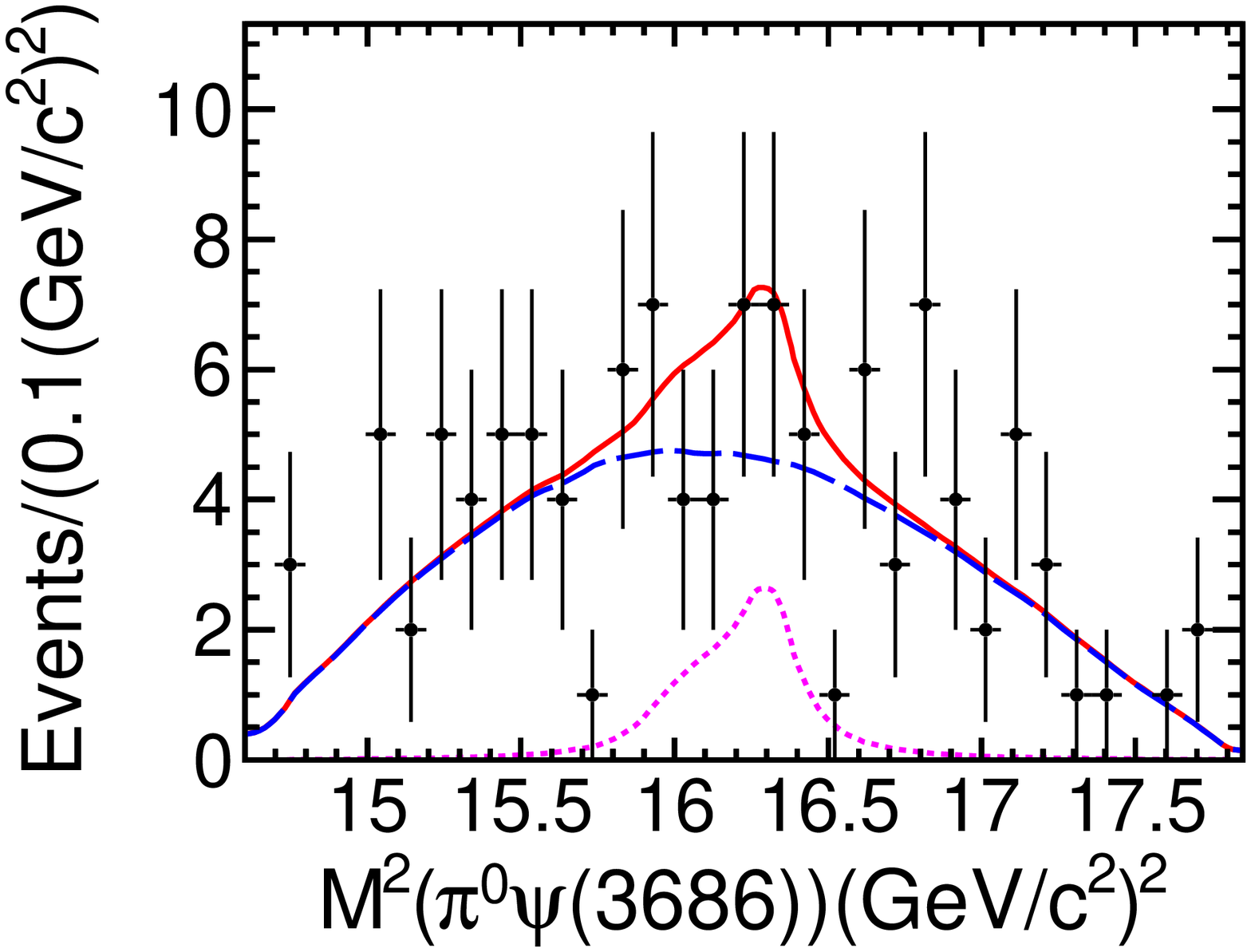}
 \end{overpic}
 \begin{overpic}[width=0.23\textwidth]{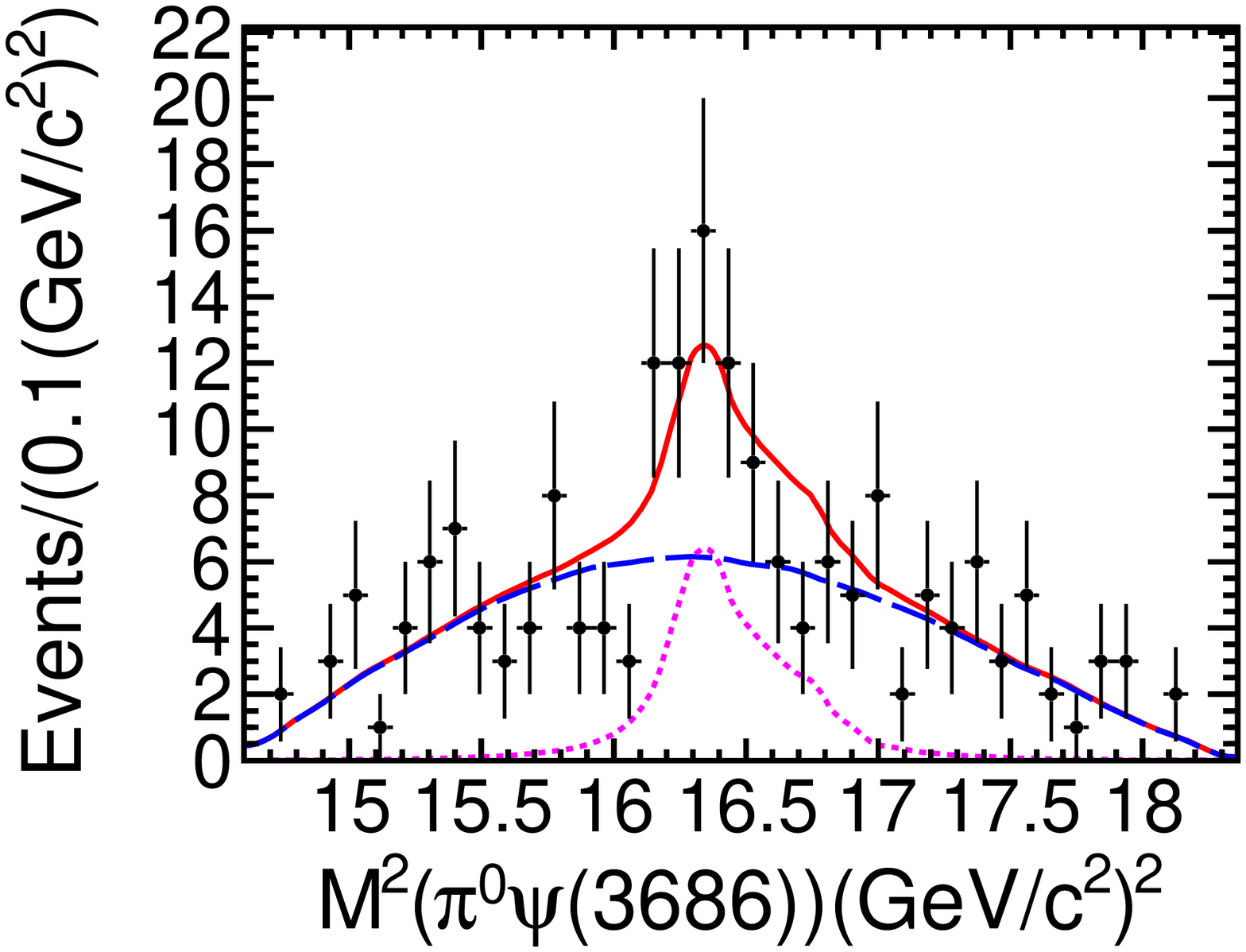}
 \end{overpic}
 \begin{overpic}[width=0.23\textwidth]{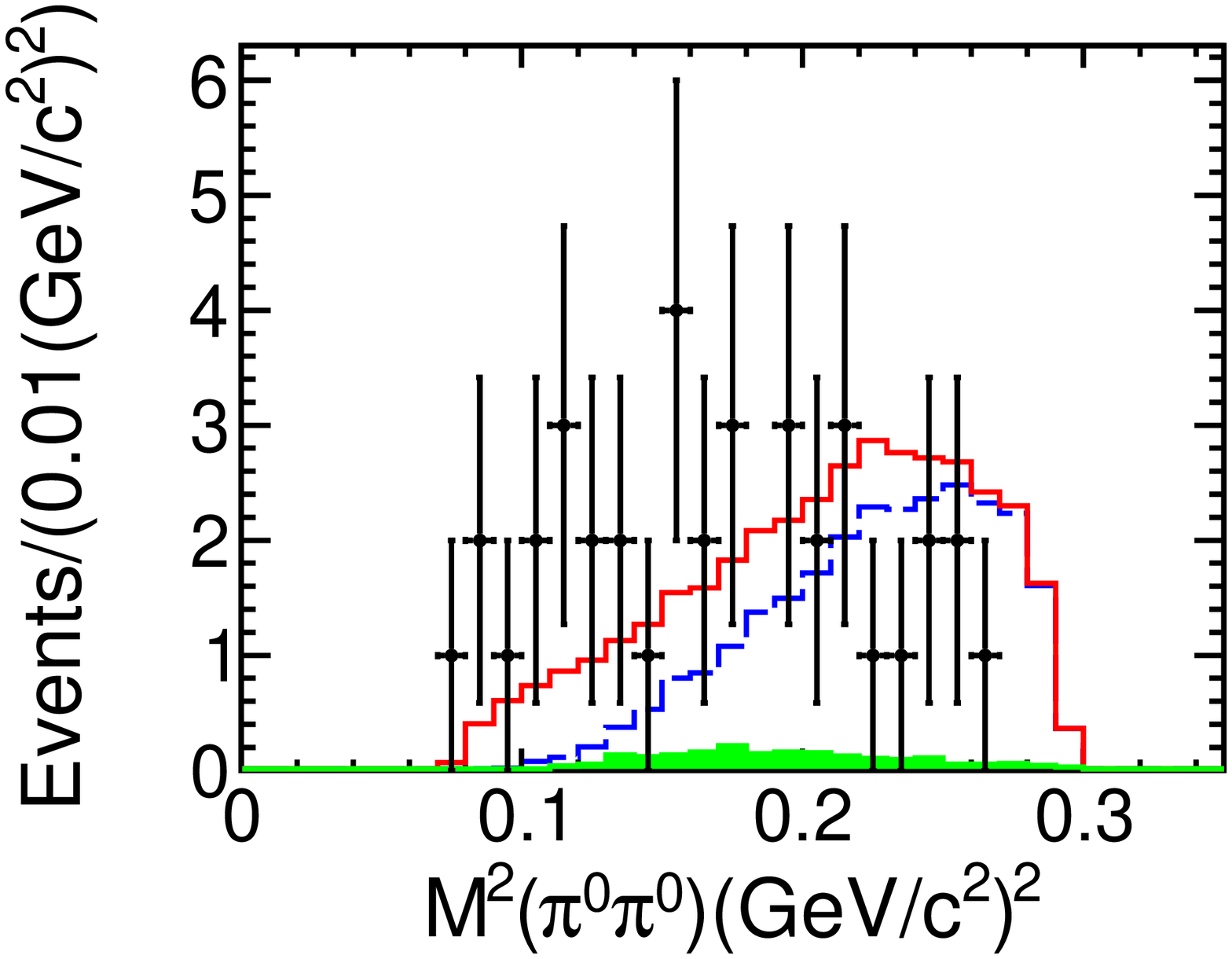}
 \end{overpic}
 \begin{overpic}[width=0.23\textwidth]{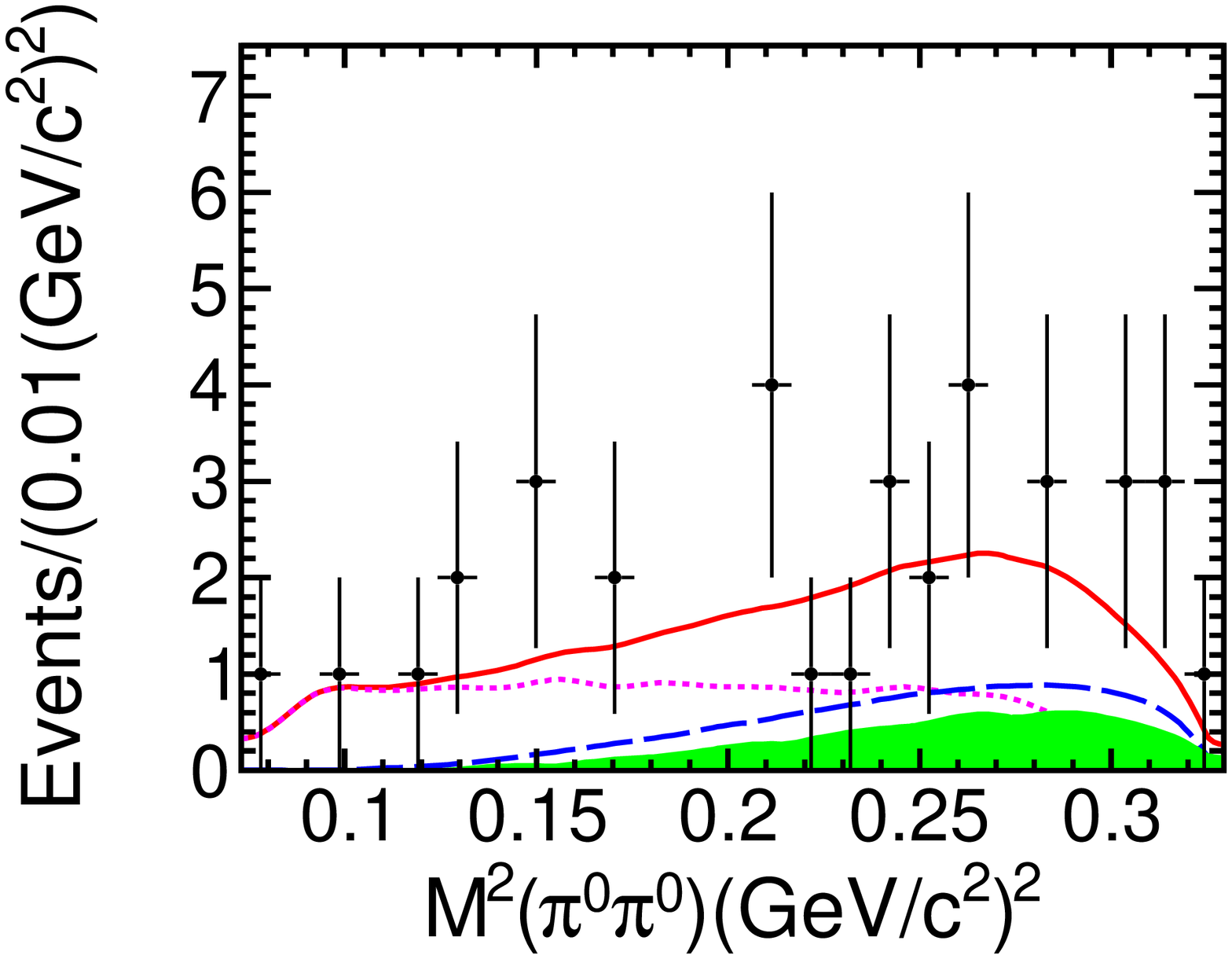}
 \end{overpic}
 \begin{overpic}[width=0.23\textwidth]{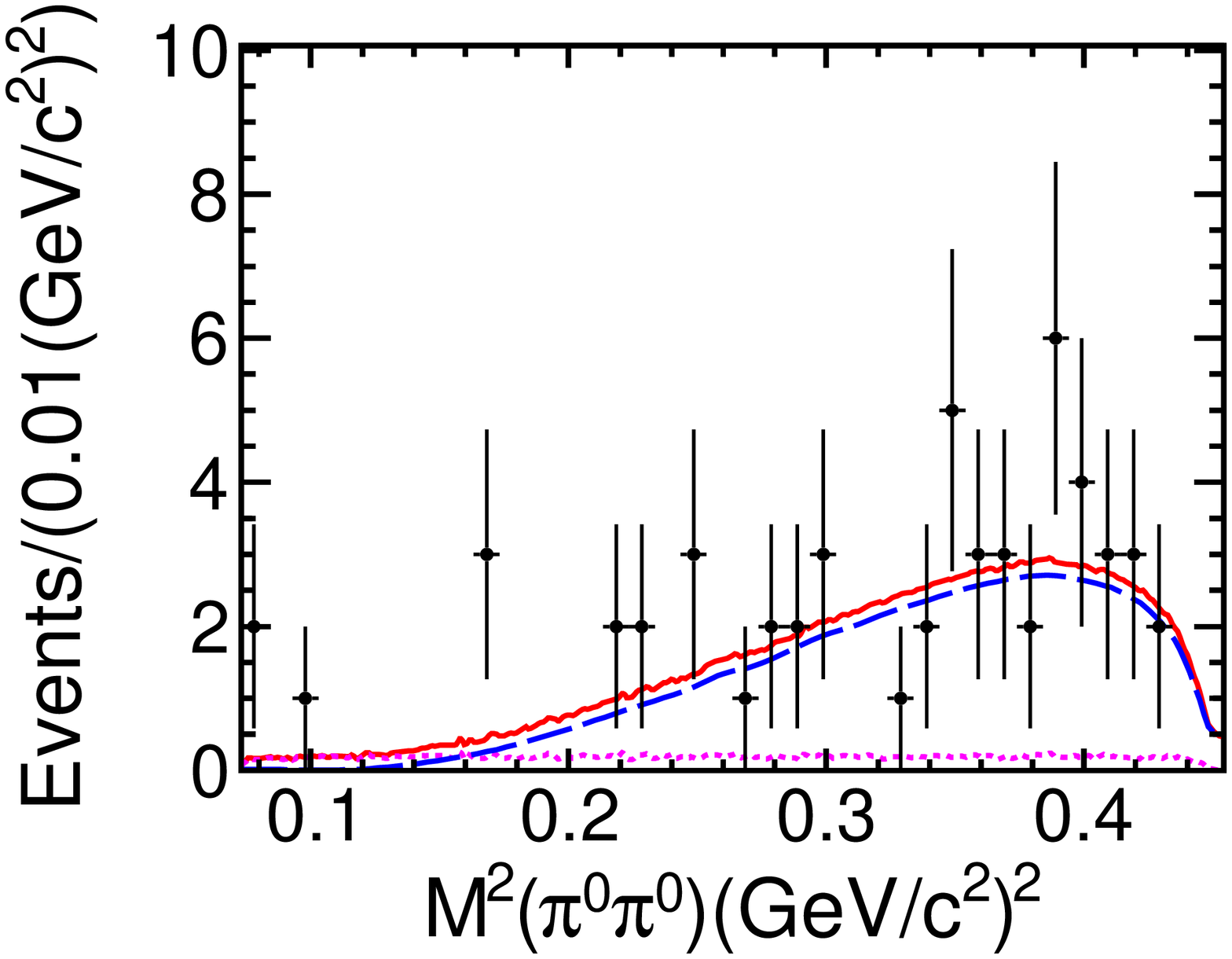}
 \end{overpic}
 \begin{overpic}[width=0.23\textwidth]{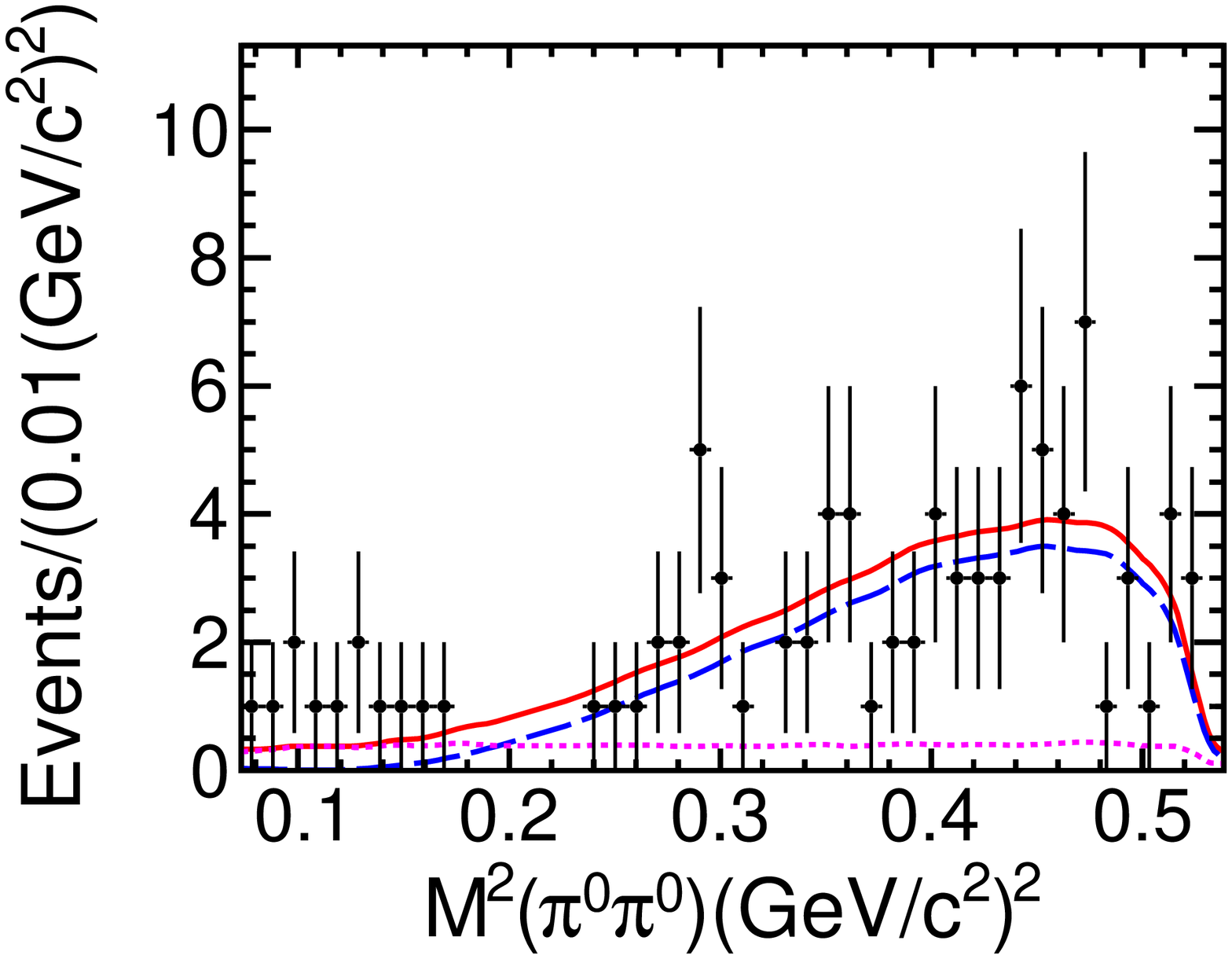}
 \end{overpic}
 \setlength{\belowcaptionskip}{-0.2cm}
 \caption{(Color online) Dalitz plots of $M^{2}(\pzpz)$ versus $M^{2}(\piz\psip)$ (top row) as well as the distributions of $M^{2}(\piz\psip)$ (middle row) and of  $M^{2}(\pzpz)$ (bottom row) for the data samples at $\sqrt{s}$ = 4.226 (column a), 4.258 (column b), 4.358 (column c), and 4.416 (column d)~$\gev$. Dots with error bars are data. For plots at $\sqrt{s}$ = 4.226~$\gev$, the red solid lines are the distributions for intermediate states, the blue dashed lines for the process $\epem\to\pzpz\psip$ simulated with the \textsc{Jpipi} model (both with an arbitrary scale). For plots at 4.258, 4.358, 4.416~$\gev$, the solid curves (red) are projections from the fits; the short dashed curves (pink) show the shapes of the intermediate states; the long dashed curves (blue) show the shapes from the direct process $\epem\to\pzpz\psip$. The green shaded histograms show the background $\epem\to\pppm\psip$ with the shape fixed to MC simulation.
}

 \label{dalitz}
\end{figure*}

Possible intermediate states in $\epem\to\pzpz\psip$ are investigated using the data samples at $\sqrt{s} = 4.226$, 4.258, 4.358, and 4.416 $\gev$.
The $\psip$ signal is extracted by selecting the events in the mass range 3.676 $<M(\pppm\lplm)<$ 3.696 $\gevcc$.
The Dalitz plots $M^{2}(\pzpz)$ versus $M^{2}(\piz\psip)$ as well as the corresponding one-dimensional distributions are shown in Fig.~\ref{dalitz}. Good agreement of these distributions with those observed in the charged mode in Ref.~\cite{Y4360Bes} is found, which confirms the variations of the kinematic behavior at different energy points and demonstrates isospin conservation.
A structure with a mass around 4040 $\mevcc$ in the $M(\piz\psip)$ spectrum is observed in the data sample at $\sqrt{s} = 4.416 $~\gev, while two bumps around 3900 and 4040 $\mevcc$ are evident in the data sample at $\sqrt{s}$ = 4.258 $\gev$.
It is worth noting that for the data sample at $\sqrt{s}$ = 4.226 $\gev$, this structure is not observed in the $M(\piz\psip)$ distribution. The behavior observed is similar with that in the charged mode~\cite{Y4360Bes}.
The dominant background is $\epem\to\pppm\psip$ as shown in Fig.~\ref{pipirecoil}. The other backgrounds are found to be negligible from the study of sideband region.

An unbinned maximum likelihood fit is performed to the Dalitz plot of $M^2(\pi^0_1\psip)$ versus $M^2(\pi^0_2\psip)$ (denoted as {\it{x}} and {\it{y}} in Eq.~(\ref{bwphse}), respectively) to determine the properties of the observed structure at $\sqrt{s} = 4.416$~\gev.
In the fit, the observed structure is assumed to be a neutral charmoniumlike state with spin-parity $1^+$, modeled with an $S$-wave Breit-Wigner function in two dimensions,
 \begin{equation}\small
 \begin{split}
    \frac{p_1\cdot q_1/c^{2}}{(x-M^{2}_R)^{2}+M_R^{2}\cdot\Gamma^{2}/c^{4}}
     + \frac{p_2\cdot q_2/c^{2}}{(y-M^{2}_R)^{2}+M_R^{2}\cdot\Gamma^{2}/c^{4}},
   \label{bwphse}
   \end{split}
 \end{equation}
taking into account the mass resolution and detection efficiency,
where, $p_{1/2}$ ($q_{1/2}$) is the momentum of the charmoniumlike state ($\psip$) in the rest frame of its mother particle, and $M_R$ and $\Gamma$ are
the mass and width of the charmoniumlike state, respectively.
The PDF of the process $\epem\to\pzpz\psip$ without an intermediate state is taken from the \textsc{Jpipi} model MC simulation.
The background is found to be negligible, and is not included in the fit.
Since the two $\piz$ mesons in the final state are experimentally indistinguishable, the fit is performed with two entries per event, and the corresponding statistical significance of the observed structure and the errors of the parameters are calculated by doubling the change of likelihood values.

The fit with a width fixed to that of the charged structure observed in $\epem\to\pppm\psip$~\cite{Y4360Bes} yields a mass of $M_R = (4038.7 \pm 6.5)$~\mevcc (consistent with that of the charged structure $M= (4032.1 \pm 2.4)$~\mevcc in Ref.~\cite{Y4360Bes}) and a statistical significance of 6.0$\sigma$ (evaluated by comparing the likelihood values with and without the charmoniumlike state included in the fit).
The fit projections on $M^2(\piz\psip)$ and $M^2(\pzpz)$ are shown in Fig.~\ref{dalitz}.
Similar to Ref.~\cite{Y4360Bes}, the fit curves are found to not match the data perfectly.
The C.L. of the fit is 19\%, estimated by toy-MC studies.
An alternative fit with free width yields a mass of $M_R= (4039.3 \pm 6.0)$ $\mevcc$,
and a width of $\Gamma= (31.9 \pm 14.8)$~\mev, which are consistent with those of the charged structure in Ref.~\cite{Y4360Bes} within the statistical uncertainties.
Another alternative fit with an additional $\zcbz$ included is performed, where the parameters of the $\zcbz$ are fixed to the weighted average values $M = (3893.6\pm3.7)$~\mevcc, $\Gamma = (31.1 \pm 7.0)$~\mev in Refs.~\cite{4020zcjpsipi0,4020zcDpi0}.
The statistical significance of the $\zcbz$ is less than 1$\sigma$.

Similar fits are carried out to the data samples at $\sqrt{s} = 4.258$ and  4.358 $\gev$, respectively, where the parameters of the charmoniumlike state are fixed to those obtained in the data sample at $\sqrt{s} = 4.416$~\gev.
In the data sample at $\sqrt{s} = 4.258$~\gev, a sizable background from $\epem\to\pppm\psip$ exists.  It is included in the fit with the
shape fixed to the MC simulation and the magnitude extracted from a fit to the $M_{\rm recoil}(\pppm)$ spectrum.
The statistical significances of the charmoniumlike structure are 3.6$\sigma$ and 4.5$\sigma$ for the data samples at $\sqrt{s} = 4.258$ and 4.358~$\gev$, respectively.
Alternative fits with additional $\zcbz$ states included are performed for the data sample at $\sqrt{s} = 4.258$~\gev.
Since both $\zcbz$ and the structure around 4040 $\mevcc$ are reflected onto each other in the $M(\piz\psip)$ spectrum, the statistical significance of $\zcbz$ is sensitive to its parameters, and is found to be 1.0$\sigma$ with the parameters above.
The fit procedure has been validated with a set of MC samples.

In summary, based on a data sample of $\epem$ collision data corresponding to 5.2~$\ifb$ at 16 c.m.~energy points between 4.009 and 4.600~$\gev$ collected with the BESIII detector, the Born cross sections for $\epem \to\pzpz\psip$ at these energy points have been measured for the first time. They are found to be half of those for $\epem\to\pppm\psip$~\cite{Y4360Bes} within uncertainties, consistent  with the expectation from isospin symmetry. The Dalitz plots of $\pzpz\psip$ are consistent with those in the $\epem\to\pppm\psip$~\cite{Y4360Bes} at all energy points. Furthermore, a structure is observed in $\pi^{0}\psip$ with a mass of ($4038.7\pm6.5$)~$\mevcc$ at $\sqrt{s}=4.416~\gev$, which confirms the structure in the charged mode. No obvious $\zcbz$ state is observed in the fit. The new observed structure may provide insight into the properties of the charged structure observed in $\epem\to\pppm\psip$ as well as the charmoniumlike $Z_c$ states observed in analogous decay modes and in charmed meson pairs. However, the fit curve does not match the data perfectly. A future larger statistics sample of data and theoretical input incorporating possible interference effects could lead to a better understanding of the structure.

The BESIII collaboration thanks the staff of BEPCII and the IHEP computing center and the supercomputing center of USTC for their strong support. This work is supported in part by National Key Basic Research Program of China under Contract No. 2015CB856700; National Natural Science Foundation of China (NSFC) under Contracts Nos. 11235011, 11335008, 11425524, 11625523, 11635010, 11375170, 11275189, 11475164, 11475169, 11605196, 11605198; the Chinese Academy of Sciences (CAS) Large-Scale Scientific Facility Program; the CAS Center for Excellence in Particle Physics (CCEPP); Joint Large-Scale Scientific Facility Funds of the NSFC and CAS under Contracts Nos. U1332201, U1532257, U1532258, U1532102; CAS Key Research Program of Frontier Sciences under Contracts Nos. QYZDJ-SSW-SLH003, QYZDJ-SSW-SLH040; 100 Talents Program of CAS; National 1000 Talents Program of China; INPAC and Shanghai Key Laboratory for Particle Physics and Cosmology; German Research Foundation DFG under Contracts Nos. Collaborative Research Center CRC 1044, FOR 2359; Istituto Nazionale di Fisica Nucleare, Italy; Koninklijke Nederlandse Akademie van Wetenschappen (KNAW) under Contract No. 530-4CDP03; Ministry of Development of Turkey under Contract No. DPT2006K-120470; National Natural Science Foundation of China (NSFC) under Contracts Nos. 11505034, 11575077; National Science and Technology fund; The Swedish Research Council; U. S. Department of Energy under Contracts Nos. DE-FG02-05ER41374, DE-SC-0010118, DE-SC-0010504, DE-SC-0012069; University of Groningen (RuG) and the Helmholtzzentrum fuer Schwerionenforschung GmbH (GSI), Darmstadt; WCU Program of National Research Foundation of Korea under Contract No. R32-2008-000-10155-0


\end{document}

%% file: def-com.tex
\newcommand{\jpsi}{J/\psi}

\newcommand{\bfg}{\begin{figure}}
\newcommand{\efg}{\end{figure}}
\newcommand{\bitm}{\begin{itemize}}
\newcommand{\eitm}{\end{itemize}}
\newcommand{\bnum}{\begin{enumerate}}
\newcommand{\enum}{\end{enumerate}}
\newcommand{\btbl}{\begin{table}}
\newcommand{\etbl}{\end{table}}
\newcommand{\btbu}{\begin{tabular}}
\newcommand{\etbu}{\end{tabular}}
\newcommand{\bcl}{\begin{center}}
\newcommand{\ecl}{\end{center}}

\newcommand{\beq}{\begin{equation}}
\newcommand{\eeq}{\end{equation}}
\newcommand{\beqr}{\begin{eqnarray}}
\newcommand{\eeqr}{\end{eqnarray}}

%% file: authors_apr2015.tex
\author{
\begin{small}
M.~Ablikim$^{1}$, M.~N.~Achasov$^{9,e}$, S. ~Ahmed$^{14}$, M.~Albrecht$^{4}$, A.~Amoroso$^{50A,50C}$, F.~F.~An$^{1}$, Q.~An$^{47,a}$, J.~Z.~Bai$^{1}$, O.~Bakina$^{24}$, R.~Baldini Ferroli$^{20A}$
% ...
\\
 \vspace{0.2cm}
 (BESIII Collaboration)\\
 \vspace{0.2cm} {\it
$^{1}$ Institute of High Energy Physics, Beijing 100049, People's Republic of China\\
% ...
$^{4}$ Bochum Ruhr-University, D-44780 Bochum, Germany\\
% ...
$^{9}$ G.I. Budker Institute of Nuclear Physics SB RAS (BINP), Novosibirsk 630090, Russia\\
% ...
$^{14}$ Helmholtz Institute Mainz, Johann-Joachim-Becher-Weg 45, D-55099 Mainz, Germany\\
% ...
$^{20}$ (A)INFN Laboratori Nazionali di Frascati, I-00044, Frascati, Italy; (B)INFN and University of Perugia, I-06100, Perugia, Italy\\
% ...
$^{24}$ Joint Institute for Nuclear Research, 141980 Dubna, Moscow region, Russia\\
% ...
$^{47}$ University of Science and Technology of China, Hefei 230026, People's Republic of China\\
% ...
$^{50}$ (A)University of Turin, I-10125, Turin, Italy; (B)University of Eastern Piedmont, I-15121, Alessandria, Italy; (C)INFN, I-10125, Turin, Italy\\
% ...
$^{a}$ Also at State Key Laboratory of Particle Detection and Electronics, Beijing 100049, Hefei 230026, People's Republic of China\\
% ...
$^{e}$ Also at the Novosibirsk State University, Novosibirsk, 630090, Russia\\
% ...
 }
\vspace{0.4cm}
}

\noaffiliation{}

%% file: draft_pi0pi0psip.bbl
\begin{thebibliography}{99}


\bibitem{Y4360BaBar} B.~Aubert {\it et al.} [BaBar Collaboration], Phys.\ Rev.\ Lett. {\bf 98}, 212001 (2007);
                     J.~P.~Lees {\it et al.} [BaBar Collaboration], Phys.\ Rev.\ D {\bf 89}, 111103 (2014).
\bibitem{Y4360Belle} X.~L.~Wang {\it et al.} [Belle Collaboration], Phys.\ Rev.\ Lett. {\bf 99}, 142002 (2007);
                     Phys.\ Rev.\ D {\bf 91}, 112007 (2015).
\bibitem{Y4360Bes}  M.~Ablikim {\it et al.} [BESIII Collaboration], Phys.\ Rev.\ D {\bf 96}, 032004 (2017).
\bibitem{NoConvention} N.~Brambilla {\it et al.}, Eur.\ Phys.\ J.\ C {\bf 71}, 1 (2011).
\bibitem{Y4260BaBar} B.~Aubert {\it et al.} [BaBar Collaboration], Phys.\ Rev.\ Lett. {\bf 95}, 142001 (2005);
                     J.~P.~Lees {\it et al.} [BaBar Collaboration], Phys.\ Rev.\ D {\bf 86}, 051102(R) (2012).
\bibitem{Y4260Cleo} Q.~He {\it et al.} [CLEO Collaboration], Phys.\ Rev.\ D {\bf 74}, 091104(R) (2006).
\bibitem{Y4260Bell} C.~Z.~Yuan {\it et al.} [Belle Collaboration], Phys.\ Rev.\ Lett. {\bf 99}, 182004 (2007);
                     Z.~Q.~Liu {\it et al.} [Belle Collaboration], Phys.\ Rev.\ Lett. {\bf 110}, 252002 (2013).
\bibitem{Y4260Bes}   M.~Ablikim {\it et al.} [BESIII Collaboration], Phys.\ Rev.\ Lett. {\bf 118}, 092001 (2017).
\bibitem{hybrid} S.~L.~Zhu, Phys.\ Lett.\ B {\bf625}, 212 (2005);
                 E.~Kou and O.~Pene, Phys.\ Lett.\ B {\bf631}, 164 (2005);
                 F.~E.~Close and P.~R.~Page, Phys.\ Lett.\ B {\bf628}, 215 (2005).
\bibitem{tetraquark} D.~Ebert, R.~N.~Faustov, and V.~O.~Galkim,  Phys.\ Lett.\ B {\bf634}, 214 (2006);
                     L.~Maiani, V.~Riquer, F.~Piccinini, and A.~D.~Polosa, Phys.\ Rev.\ D {\bf 72},
                     031502(R) (2005);T.~W.~Chiu, and T.~H.~Hsieh [TWQCD Collaboration],  Phys.\ Rev.\ D {\bf 73}, 094510 (2006).
\bibitem{molecule}   X.~Liu, X.~Q.~Zeng, and X.~Q.~Li, Phys.\ Rev.\ D {\bf 72}, 054023(R) (2005);
                     C.~F.~Qiao, Phys.\ Lett.\ B {\bf639}, 263 (2006);
                     C.~Z.~Yuan, P.~Wang, and X.~H.~Mo, Phys.\ Lett.\ B {\bf634}, 399 (2006).
\bibitem{ZC3900BES}  M.~Ablikim {\it et al.} [BESIII Collaboration], Phys.\ Rev.\ Lett. {\bf 110}, 252001 (2013).
\bibitem{ZC3900CLEO} T.~Xiao {\it et al.}, Phys.\ Lett.\ B {\bf 727}, 366 (2013).
\bibitem{ZC4020BES} M.~Ablikim {\it et al.} [BESIII Collaboration], Phys.\ Rev.\ Lett. {\bf 111}, 242001 (2013).
\bibitem{ZC3885BES} M.~Ablikim {\it et al.} [BESIII Collaboration], Phys.\ Rev.\ Lett. {\bf 112}, 022001 (2014).
\bibitem{ZC3885BES2} M.~Ablikim {\it et al.} [BESIII Collaboration], Phys.\ Rev.\ D {\bf 92}, 092006 (2015).
\bibitem{ZC4025BES} M.~Ablikim {\it et al.} [BESIII Collaboration], Phys.\ Rev.\ Lett. {\bf 112}, 132001 (2014).
\bibitem{zb} A.~Bondar {\it et al.} [Belle Collaboration], Phys.\ Rev.\ Lett. {\bf 108},122001(2012).
\bibitem{exotics}  H.~X.~Chen {\it et al.}, Phys.\ Rept.\  {\bf 639}, 1 (2016).
\bibitem{4020zchc}  M.~Ablikim {\it et al.} [BESIII Collaboration], Phys.\ Rev.\ Lett. {\bf 113}, 212002 (2014).
\bibitem{4020zcDpi0}M.~Ablikim {\it et al.} [BESIII Collaboration], Phys.\ Rev.\ Lett. {\bf 115}, 222002 (2015).
\bibitem{4020zcDDs}  M.~Ablikim {\it et al.} [BESIII Collaboration], Phys.\ Rev.\ Lett. {\bf 115}, 182002 (2015).
\bibitem{4020zcjpsipi0}  M.~Ablikim {\it et al.} [BESIII Collaboration], Phys.\ Rev.\ Lett. {\bf 112}, 132001 (2014).
\bibitem{luminosity} M.~Ablikim {\it et al.} [BESIII Collaboration], Chin.\ Phys.\ C {\bf 39}, 093001 (2015).
\bibitem{energycorr} M.~Ablikim {\it et al.} [BESIII Collaboration], Chin.\ Phys.\ C {\bf 40}, 063001 (2016).
\bibitem{besint} M.~Ablikim {\it et al.} [BESIII Collaboration], Nucl.\ Instrum.\ Meth.\ A {\bf 614}, 3 (2010).
\bibitem{geant4} S.~Agostinelli {\it et al.} [{\sc Geant4} Collaboration], Nucl.\ Instrum.\ Meth. A {\bf 506}, 250 (2003).
\bibitem{Deng} Z.~Y.~Deng \textit{et~al.}, HEP\&NP {\bf 30}, 371 (2006).
\bibitem{kkmc1} S.~Jadach, B.~F.~L.~Ward, and Z.~Was, Comput. Phys. Commun. {\bf 130}, 260 (2000); Phys.\ Rev.\ D {\bf 63}, 113009 (2001);
                A.~Fotopoulos and M.~Tsulaia, Int.\ J.\ Mod.\ Phys.\ A {\bf 24}, S1 (2009).
\bibitem{jpipi} T.~Mannel and R.~Urech, Z.\ Phys.\ C {\bf 73}, 541 (1997).
\bibitem{EVTGEN} R.~G.~Ping, Chin.\ Phys.\ C {\bf 32}, 599 (2008);
                D.~J.~Lange, Nucl.\ Instrum.\ Meth.\ A {\bf 462}, 152 (2001).
\bibitem{ISR} E. A.~Kuraev and V.~S.~Fadin, Sov. J.\ Nucl.\ Phys. {\bf 41}, 466 (1985).
\bibitem{VR} S.~Actis {\it et al.} Eur.\ Phys.\ J.\ C {\bf 66}, 585 (2010).
\bibitem{PDG} C.~Patrignani {\it et al.} [Particle Data Group], Chin.\ Phys.\ C, {\bf 40}, 100001  (2016).
\bibitem{supp} See Supplemental Material at [URL wil be inserted by the publisher] for a summary of the luminosity, signal events, efficiency and Born cross section at each energy point, and systematic errors.
\bibitem{Rolke} W. A.~Rolke, A. M.~Lopez and J. Conrad, Nucl.\ Instr.\ Methods Phys.\ Res.\ Sect. A {\bf 551}, 493 (2005).
\bibitem{trackerror} M.~Ablikim {\it et al.} [BESIII Collaboration],  Phys.\ Rev.\ Lett. {\bf 112}, 022001 (2014); Phys.\ Rev.\ D {\bf 83}, 112005 (2011).
\bibitem{photonerror} M.~Ablikim {\it et al.} [BESIII Collaboration], Phys.\ Rev.\ D {\bf 81}, 052005 (2010).
\bibitem{helixsys} M.~Ablikim {\it et al.} [BESIII Collaboration], Phys.\ Rev.\ D {\bf 87}, 012002 (2013).

\end{thebibliography}
